\documentclass[aps,prl,reprint,showpacs,floatfix]{revtex4-1}

\usepackage{amsmath}
\usepackage{slashed}
\usepackage{txfonts}
\usepackage{microtype}
\usepackage{graphicx}
\usepackage[breaklinks=true]{hyperref}
\usepackage{color}

\def\sout{\bgroup\markoverwith
{\textcolor{red}{\rule[0.5ex]{2pt}{0.5pt}}}\ULon}
\def\be{\begin{equation}}
\def\ee{\end{equation}}
\def\bes{\begin{equation*}}
\def\ees{\end{equation*}}
\def\bea{\begin{eqnarray}}
\def\eea{\end{eqnarray}}
\def\beas{\begin{eqnarray*}}
\def\eeas{\end{eqnarray*}}
\def\bal#1\eal{\begin{align}#1\end{align}}
\def\bals#1\eals{\begin{align*}#1\end{align*}}
\newcommand{\bra}[1]{\langle #1|}
\newcommand{\ket}[1]{|#1\rangle}

\renewcommand{\vec}{\vectorsym}

\renewcommand*{\vec}[1]{\boldsymbol{#1}}

\graphicspath{{figures/}}

\usepackage[normalem]{ulem}

\begin{document}




\title{Quantum groups as hidden symmetries of quantum impurities}

\author{E. Yakaboylu}
\email{enderalp.yakaboylu@ist.ac.at}
\author{M. Shkolnikov}
\email{mikhail.shkolnikov@ist.ac.at}
\author{M. Lemeshko}
\email{mikhail.lemeshko@ist.ac.at}

\affiliation{IST Austria (Institute of Science and Technology Austria), Am Campus 1, 3400 Klosterneuburg, Austria}

\date{\today}

\begin{abstract}

We present an approach to interacting quantum many-body systems based on the notion of quantum groups, also known as $q$-deformed Lie algebras. In particular, we show that if the symmetry of a free quantum particle corresponds to a Lie group $G$, in the presence of a  many-body environment this particle can be described by a deformed group, $G_q$. Crucially, the single deformation parameter, $q$, contains all the information about the many-particle interactions in the system. We exemplify our approach by considering a quantum rotor interacting with a bath of bosons, and demonstrate that extracting the value of $q$ from closed-form solutions in the perturbative regime allows one to predict the behavior of the system for arbitrary values of the impurity-bath coupling strength, in good agreement with non-perturbative calculations. Furthermore, the value of the deformation parameter allows to predict at which coupling strengths rotor-bath interactions result in a formation of a stable quasiparticle. The approach based on quantum groups does not only allow for a drastic simplification of impurity problems, but also provides valuable insights into hidden symmetries of interacting many-particle systems.

\end{abstract}

\maketitle

The development of physics is accompanied by the study of symmetries possessed by natural phenomena, and hence by the study of corresponding groups and their representations. Perhaps the most iconic example is the symmetry group of special relativity, which is known as the Poincar\'{e} group, whose irreducible representations clasify elementary particles~\cite{weinberg1995quantum}. Other important examples include gauge symmetry leading to the Standard Model, point groups that describe symmetries of crystal lattices in solid state physics, and conformal symmetry which lies at the base of the string theory and explains several critical phenomena~\cite{cornwell1997group,pelissetto2002critical}. 

During the last decades there has been a great interest in the study of quantum groups, which correspond to deformations of the conventional Lie algebras~\cite{drinfeld1987proceedings,jimbo1986q,jimbo1990yang,kulish1981quantum}. From a mathematician's perspective, quantum groups are Hopf algebras, which possess a coproduct, a counit, and an antipode, in addition to the regular structures of an algebra~\cite{abe1977hopf}. In physics, quantum groups have been first applied to solve the quantum Yang-Baxter equation~\cite{jimbo1990yang}, and, over the years, found several  applications to spin chains~\cite{batchelor1990q}, anyons~\cite{lerda1993anyons,ubriaco1997anyonic}, quantum
optics~\cite{buvzek1992jaynes,chaichian1990quantum}, and rotational and vibrational molecular spectra~\cite{esteve1992q,raychev1995quantum} (for further details see Ref.~\cite{bonatsos1999quantum} and references therein).

In this Letter, we show that quantum groups can be used to drastically simplify the problems of quantum many-particle physics. In particular, we demonstrate that if the symmetry of an isolated quantum particle corresponds to a Lie group $G$, in the presence of a many-body environment this particle  can be described by a ``deformed'' quantum group, $G_q$. Crucially, all the interactions of the quantum impurity with the surrounding many-body bath are contained in the so-called deformation parameter, $q$.

For the sake of concreteness, let us  consider a linear rigid rotor interacting with a bath of bosons. An isolated rotor can be described by the SO(3) group of rotations in three-dimensional space~\cite{Varshalovich}. Our claim is that once the rotor is immersed in a many-particle environment, the resulting composite object can be described by the quantum group, SO$_q$(3). Let us start from the Hamiltonian describing the interactions between a quantum rotor and the surrounding many-particle bath of bosons,
\be
\label{angulon_ham}
\hat H = \hat H_\text{rotor}  + \hat H_\text{bath} + \hat H_\text{int} \, .
\ee
Here the first term, $\hat H_\text{rotor} = B \, \vec{\hat{J}}^2$, represents the rotational kinetic energy of a linear rigid rotor with the rotational constant $B$. The second term, $\hat H_\text{bath} = \sum_{k \lambda \mu} \omega (k) \, \hat{b}^\dagger_{k \lambda \mu} \hat{b}_{k \lambda \mu}$, with $\sum_k \equiv \int d k$, corresponds to the kinetic energy of the bosons parametrised by the dispersion relation, $\omega(k)$. Here $\hat{b}^\dagger_{k \lambda \mu} $ and $\hat{b}_{k \lambda \mu}$ are the bosonic creation and annihilation operators cast in the angular momentum representation, with $k$, $\lambda$, and $\mu$ labeling the bosonic linear momentum, angular momentum, and its projection on the laboratory-frame $z$-axis, respectively. The last term of Eq.~\eqref{angulon_ham} describes the interaction of the impurity with the bosonic bath~\cite{Lemeshko_2015},
\be
\label{Hint}
\hat H_\text{int} =  \sum_{k \lambda \mu} U_{\lambda} (k) \left[ Y^{*}_{\lambda \mu}(\hat{\Omega}) \hat{b}^\dagger_{k \lambda \mu} + Y_{\lambda \mu}(\hat{\Omega}) \hat{b}_{k \lambda \mu}  \right] \, ,
\ee
where $Y_{\lambda \mu}(\hat{\Omega})$ are the spherical harmonic operators~\cite{Varshalovich} that depend on the impurity orientation in the laboratory frame, $\hat \Omega \equiv (\hat \theta, \hat \phi)$, and $U_{\lambda} (k)$ is the angular-momentum-dependent coupling strength. The Hamiltonian of the form~\eqref{angulon_ham} was shown to describe the so-called angulon quasiparticle that has been studied in the context of experiments on molecules in superfluid helium nanodroplets~\cite{PhysRevX.6.011012,lemeshko2016quasiparticle,Shepperson16,Cherepanov}.

In the absence of a many-particle bath, the Hamiltonian~\eqref{angulon_ham} reduces to $\hat H_\text{rotor}$, which is nothing else but the SO(3) Casimir operator, $\vec{\hat{J}}^2$, that commutes with all the elements of the corresponding Lie algebra. Therefore, the eigenvalues of $\hat H_\text{rotor}  $ simply follow from the Casimir values of the Lie group $\text{SO} (3)$, $E_j = B \, j (j+1)$. Many-particle interactions, Eq.~\eqref{Hint}, result in ``dressing'' of  the rotor by a cloud of bosonic excitations, which  induce renormalization of the rotor's rotational constant, $B$, to some value $B^\ast <B$. Calculating $B^*$ numerically can be extremely challenging for it involves addition of a macroscopic number of angular momenta~\cite{Lemeshko_2016_book}. In what follows, we show that such a calculation can be drastically simplified by casting the problem in the language of quantum groups. We aim to show that it is possible to find a deformed quantum algebra, $\text{SO}_q (3)$, such that its deformation, $q$, describes the rotor-bath interactions of Eq.~\eqref{Hint}.

The renormalized rotational constant, in analogy to the polaron effective mass~\cite{Devreese15}, is given by the second-order finite difference,
\be
\label{eqBst}
B^* =\frac{1}{2}\sum_{j=0}^2 (-1)^j \binom{2}{j} E_{2-j} \, ,
\ee   
where $E_{j}$ is the eigenenergy of $\hat H$ corresponding to total angular momentum $j$. In the weak-coupling regime, where the interaction term, $\hat H_\text{int}$, is small compared to $\hat H_\text{rotor}$, the impurity energy can be calculated within the perturbation theory. Up to the second order in $U_\lambda (k)$, the perturbed energy is written as
\bal
\label{eqEj}
\nonumber E_j^\text{ang} & = B j (j+1) + \sum_{j'm' k \lambda \mu}\frac{\left| \bra{j'm'} \bra{0} \hat b_{k \lambda \mu} \hat H_\text{int} \ket{jm} \ket{0} \right|^2}{B j(j+1) - B j'(j'+1) - \omega(k)} \\
& + \mathcal{O}(U_\lambda (k)^4)\, ,
\eal
with $\ket{0}$ being the vacuum state of the bath. From \eqref{eqBst} and \eqref{eqEj}, we obtain:
\bal
\label{bs_weak_angulon}
\nonumber B^* &= B - \frac{1}{2}\sum_{j=0}^2  \sum_{k \lambda j'}  \binom{2}{j} \frac{(-1)^{j} V_\lambda (k)^2 \left[ C_{2-j 0, \lambda 0}^{j' 0} \right]^2}{B j'(j'+1) + \omega(k) - B (2-j)(3-j)} \\
& + \mathcal{O}(U_\lambda(k)^4) \, ,
\eal
where $V_\lambda(k) = U_\lambda (k) \sqrt{(2 \lambda +1)/(4 \pi)}$.

Eq.~(\ref{bs_weak_angulon}) shows how a many-particle environment deforms the rotational constant of a linear rotor, $B \to B^\ast$. Our goal is to show that such a deformation can also be obtained within the  quantum group $\text{SO}_q (3)$.  In what follows, we first briefly describe quantum groups and introduce elementary tools for $\text{SO}_q (3)$ needed for the description of a quantum rotor immersed in a many-particle bath, and then identify the deformation parameter, $q$, in terms of the bath degrees of freedom.

First of all, quantum groups are Hopf algebras which are deformations of Lie groups. In addition to associative product and a unit element, Hopf algebras possess a coproduct, a counit, and an antipode. These operations are responsible for tensor product on representations, trivial one-dimensional representation, and duality,  respectively. The algebra of functions on a classical group is a commutative Hopf algebra with coproduct inherited from group multiplication, counit from unit, and antipode from group inversion. Its non-commutative deformations are seen as functions on a quantum group.  While in the present Letter we do not explicitly use any of these exotic structures, their importance becomes apparent if one considers a system of several impurities.

For a given Lie algebra, its universal enveloping algebra has cocommutative  coproduct given by $x\mapsto 1\otimes x+x\otimes 1$ for any $x$ in the Lie algebra. Essentially, it is the dual Hopf algebra to the algebra of functions on a corresponding Lie group. The quantum group SO$_q$(3), in turn, is a non-cocommutative deformation of the universal enveloping algebra of the Lie algebra of SO(3). As a unital associative algebra, it is generated by $\hat{J}_z^q$, $\hat{J}_{+}^q$, $\hat{J}_{-}^q$ satisfying the following commutation relations:
\be
[\hat{J}_z^q, \hat{J}_{\pm}^q] = \pm \hat{J}_{\pm}^q \quad [\hat{J}_{+}^q, \hat{J}_{-}^q] = [2\hat{J}_z^q]_q \, ,
\ee
where the square bracket implies 
\be
[\hat{A}]_q = \frac{q^{\hat{A}} - q^{-\hat{A}}}{q-q^{-1}} \, ,
\ee
such that in the limit of $q\to 1$ one recovers $[\hat{A}]_q \to \hat{A}$.

The corresponding Casimir operator is given by $ \mathcal{\hat{C}}_q = \hat{J}_{-}^q \hat{J}_{+}^q + [\hat{J}_z^q]_q [\hat{J}_z^q + 1]_q $, and hence the Hamiltonian of an object that obeys symmetries of the quantum group SO$_q$(3) can be written as \be
\label{q_ham}
\hat H_q = B \, \mathcal{\hat{C}}_q \, .
\ee
The eigenvalues of the deformed Hamiltonian~(\ref{q_ham}) are given by $ E_j^q = B \, [j]_q[j+1]_q $. Furthermore, because the eigenvalues have to be real, the deformation parameter $q$ can be written in the form of $q = e^{i \tau}$, with $\tau$ being either a real or an imaginary number. This allows us to write
\be
\label{quant_casimir}
E_j^q = B \, \frac{\sin[\tau j] \sin[\tau (j+1)]}{\sin^2 (\tau)} \, .
\ee

From the energy~(\ref{quant_casimir}), one can calculate the renormalized rotational constant as
\be
\label{bstar}
B^* = B \cos(3 \tau) \, .
\ee
For small values of deformation, this gives $B^* = B (1 - 9 \tau^2/2 + \mathcal{O}(\tau^4))$. Then, by matching the latter with Eq.~(\ref{bs_weak_angulon}), where we have assumed weak coupling between the impurity and the bath, we obtain:
\be
\label{deformation_parameter}
\tau = \left( \frac{1}{9 B} \sum_{k \lambda j' j}  \binom{2}{j} \frac{(-1)^{j} V_\lambda (k)^2 \left[ C_{2-j 0, \lambda 0}^{j' 0} \right]^2}{B j'(j'+1) + \omega(k) - B (2-j)(3-j)} \right)^{1/2} \, ,
\ee
which is the main result of the paper. Let us now demonstrate that $B^*$ given by Eqs.~(\ref{bstar}) and~(\ref{deformation_parameter}) is valid not only for weak interactions, but for \textit{arbitrary} values of the impurity-bath coupling strength. 

To show that, as the first step we calculate renormalization of the rotational constant in the opposite, strong-coupling regime. There, the renormalized rotational constant can be calculated by expanding Eq.~(\ref{bstar}) for small values of $B$ as follows:
\be
\label{bs_strong_qg}
B^* =  B - 2B^2 \sum_{k \lambda} \frac{ V_\lambda(k)^2}{\omega(k)^3} \lambda (\lambda+1)  + \mathcal{O}(B^3) \, ,
\ee
where we used that $\sum_{j' j}  \binom{2}{j} (-1)^{j}\left[ C_{2-j 0, \lambda 0}^{j' 0} \right]^2 = 0 $ for $\forall \lambda$.

Let us compare the renormalized rotational constant~(\ref{bs_strong_qg}) with the result predicted by  the angulon theory. For this purpose, we first rewrite the Hamiltonian~\eqref{angulon_ham} in the co-rotating frame~\cite{PhysRevX.6.011012}
\bal
\label{angulon_ham_1}
\nonumber \hat H_\text{ang}' & = \hat S^{-1} \hat H_\text{ang} \hat S = B \, \vec{\hat{J}'}^2 + \sum_{k \lambda \mu} \omega_\lambda (k) \, \hat{b}^\dagger_{k \lambda \mu} \hat{b}_{k \lambda \mu} \\ 
& + \sum_{k \lambda} V_{\lambda} (k) \left[\hat{b}^\dagger_{k \lambda 0} +  \hat{b}_{k \lambda 0}  \right] + \hat H'_\text{int} \, ,
\eal
where $\hat S = e^{-i \varphi \hat \Lambda_z} e^{-i \theta \hat \Lambda_y} e^{-i \gamma \hat \Lambda_z} $, $\vec{\hat{J}}'$ is the anomalous angular momentum operator in the body-fixed frame, and $\omega_\lambda (k) = \omega (k) + B \lambda (\lambda +1)$. The interaction term, on the other hand, is given by
\be
\hat H'_\text{int} = - 2 B \, \vec{\hat{J}'} \cdot \hat{\vec{\Lambda}} + B \, \hat \Gamma \, .
\ee
Here $\hat{\vec{\Lambda}} = \sum_{k \lambda \mu \nu} \vec{\sigma}^\lambda_{\mu \nu} \hat{b}^\dagger_{k \lambda \mu} \hat{b}_{k \lambda \nu} $ is the angular momentum of the bath with $\vec{\sigma}^\lambda_{\mu \nu}$ being the angular-momentum-$\lambda$ representation of SO(3). Furthermore, while the first term $\vec{\hat{J}'} \cdot \hat{\vec{\Lambda}}$ defines the impurity-bath interaction, the last term $\hat \Gamma = \sum_{k \lambda \mu \nu} \sum_{k' \lambda' \mu' \nu'} \vec{\sigma}^\lambda_{\mu \nu} \cdot \vec{\sigma}^{\lambda'}_{\mu' \nu'} \hat{b}^\dagger_{k \lambda \mu} \hat{b}^\dagger_{k' \lambda' \mu'} \hat{b}_{k \lambda \nu} \hat{b}_{k' \lambda' \nu'}$ is the effective phonon-phonon interaction in the rotating frame.

\begin{figure}
  \centering
  \includegraphics[width=0.9\linewidth]{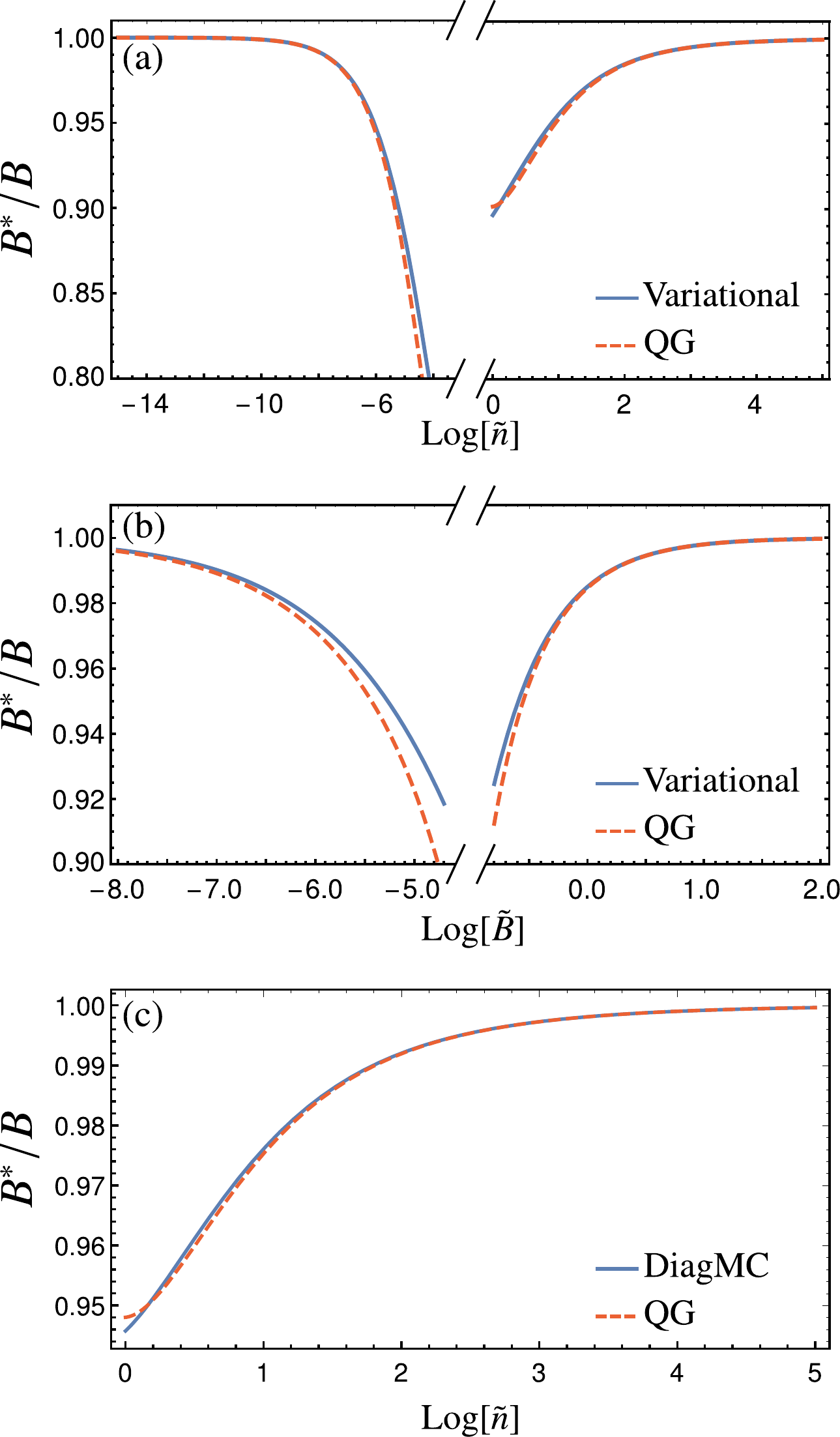}
 \caption{Comparison of the renormalized rotational constant, $B^*/B$, obtained within the quantum group (QG) approach and using the variational method of Ref.~\cite{PhysRevX.6.011012} (a) as a function of dimensionless bath density, $\tilde{n}$, for the parameters of Ref.~\cite{Lemeshko_2015}, and (b) as a function of dimensionless rotational constant $\tilde{B}$ for the parameters of Ref.~\cite{PhysRevX.6.011012}. (c) The comparison with the Diagrammatic Monte Carlo (DiagMC) approach as a function of $\tilde{n}$ for the parameters used in Ref.~\cite{bighin2018diagrammatic}. See the text.}
 \label{variational}
\end{figure}

For small values of the rotational constant $B$, the interaction Hamiltonian~$\hat H'_\text{int} $ can be treated as a perturbation, and the corresponding energy can be calculated within the perturbation theory. The unperturbed eigenstate can be written as $ \ket{jm0} \otimes U \ket{0} $, where $\ket{jmn}$ is the eigenstate of $\vec{\hat{J}'}^2$, and $\hat U = \exp\left( - \sum_{k \lambda} (\hat{b}^\dagger_{k \lambda 0} -  \hat{b}_{k \lambda 0}) V_\lambda(k)/(\omega_\lambda (k)) \right)$ diagonalizes the unperturbed bosonic Hamiltonian. The unperturbed eigenvalue is ${E'}_j^{\text{ang} \,(0)} =  B j(j+1) - \varepsilon_0 $, where the so-called deformation energy is given by $\varepsilon_0 = \sum_{k \lambda} V_\lambda(k)^2 / \omega_\lambda (k) $.

Up to the second order in $B$, the perturbed energy reads:
\bal
\nonumber & {E'}_j^{\text{ang}}  = B j(j+1) - \varepsilon_0   - \sum_{k \lambda \mu n} \frac{\left| \bra{jmn} \bra{0} \hat b_{k \lambda \mu} \hat U^{-1} \hat H'_\text{int} \hat U \ket{0} \ket{jm 0} \right|^2}{\omega_\lambda (k)} \\
 &  - \sum_{k' \lambda' \mu'} \sum_{k \lambda \mu n} \frac{\left| \bra{jmn} \bra{0} \hat b_{k \lambda \mu} \hat b_{k' \lambda' \mu'} \hat U^{-1} \hat H'_\text{int} \hat U \ket{0} \ket{jm 0} \right|^2}{\omega_\lambda (k)+\omega_{\lambda'} (k')}  + \mathcal{O}(B^3)\, .
\eal
Note that in contrast to the weak-coupling regime, the perturbed energy of the strong-coupling approach also includes two-phonon states. The corresponding renormalized rotational constant is then given by:
\be
\label{bs_strong_angulon}
B^* = B - 2B^2 \sum_{k \lambda} \frac{V_\lambda (k)^2}{\omega (k)^3} \lambda(\lambda+1) + \mathcal{O}(B^3) \, ,
\ee
which coincides with Eq.~(\ref{bs_strong_qg}) exactly at second order. 

We would like to emphasize that the result of Eq.~(\ref{bs_strong_angulon}) cannot be obtained directly from the weak-coupling perturbative result, Eq.~(\ref{bs_weak_angulon}), which gives $B^* = B - B^2 \left( 2 \sum_{k \lambda} V_\lambda (k)^2 \lambda(\lambda+1)/\omega (k)^3 + \mathcal{O}(U_\lambda (k)^4) \right) + \mathcal{O}(B^3)$. Therefore, the deformation parameter $\tau$ connects these two opposite expansions in a consistent way. We further note that one cannot deduce $\tau$ starting from Eq.~(\ref{bs_strong_angulon}), since $\tau$ describes the deformation of a quantum rotor by a many-body bath and not the other way around.

The analytical agreement between the quantum group approach and the perturbation theory in the strong-coupling regime is the first signature of that the rigid rotor dressed by bosons (or the ``angulon''~\cite{Lemeshko_2016_book}) can be described within the quantum group SO$_q$(3). For further justifications, we aim to go beyond perturbative techniques and compare Eq.~(\ref{bstar}) with  non-perturbative results obtained within various many-body techniques.  For this purpose, we consider a bath with the Bogoliubov dispersion relation, $\omega (k) = \sqrt{\epsilon (k) (\epsilon (k) + 2 g_{\text{bb}} n)}$~\cite{Pitaevskii2016}, where $\epsilon(k) = k^2/2m_b$, with $m_b$ being the boson mass,
$g_\text{bb}$ is the boson-boson contact interaction, and $n$ is the boson particle density, and choose the impurity-boson interaction of the fom $U_\lambda (k) = \sqrt{8 n k^2 \epsilon (k) /(\omega (k) (2\lambda+1))} \int dr r^2 V_\lambda (r) j_\lambda (k r)$, where the coupling is modeled by using Gaussian functions, $V_\lambda (r) =  u_\lambda (2\pi)^{-3/2} e^{-r^2/(2 r_\lambda^2)}$, and focus on the leading orders, $\lambda =0,1$.

First, we compare the quantum group approach with a variational method. In Ref.~\cite{PhysRevX.6.011012}, it was shown that the angulon problem can be solved using a variational ansatz based on single-phonon excitations on top of a bosonic coherent state. In Fig.~\ref{variational} (a), we show comparison
of Eq.~(\ref{bstar}) and the renormalized rotational constant obtained within the variational method for the parameters of Ref.~\cite{Lemeshko_2015}.  Namely, we set the parameters to $g_{\text{bb}} = 4\pi a_{\text{bb}}/m_b$, with $a_{\text{bb}} = 3.3 /\sqrt{m_b B}$, $u_0 = 1.75 u_1 = 218 B$, and $r_0 = r_1 = 1.5 /\sqrt{m_b B}$, and present the results as a function of the dimensionless bath density, $\tilde{n} = n (m_b B)^{3/2} $. One can see that a good agreement is obtained. In Fig.~\ref{variational} (b), we compare the quantum group approach with the variational method for the paremeters given in Ref.~\cite{PhysRevX.6.011012} ($g_\text{bb} = 418 (m_b^2 u_0)^{-1/2}$, $u_1 = 5 u_0$, and $r_0 = r_1 = 15 (m_b u_0)^{-1/2}$). There, we present the results  as a function of the dimensionless rotational constant, $\tilde{B} = B/u_0$, and obtain a similarly good agreement.  As the next step, we go beyond the variational method, and compare the quantum group approach with the diagrammatic Monte Carlo (DiagMC) technique~\cite{bighin2018diagrammatic}, which is applicable at arbitrary coupling. Fig.~\ref{variational}~(c), plotted for the parameters of  Ref.~\cite{bighin2018diagrammatic} ($g_{\text{bb}} = 4\pi a_{\text{bb}}/m_b$ with $a_{\text{bb}} = 3.3 /\sqrt{m_b B}$, $u_0 = 3.33 u_1 = 300 B$, and $r_0 = r_1 = 1.5 /\sqrt{m_b B}$), reveals an unprecedented agreement, which indicates that the quantum group approach is a promising method to calculate $B^*$. We note that the results are not plotted for the range at intermediate coupling where the quasiparticle picture fails, see the discussion below. 

The eigenvalues of the quantum Casimir operator, Eq.~(\ref{quant_casimir}), can be expanded in terms of the classical Casimir values, $E^q_j = B \sum_n a_n j^n (j+1)^n$. Furthermore, when the deformation parameter $\tau$ is real, the expansion turns into an alternating series~\cite{raychev1995quantum}. In fact, when the series has alternating signs, the expansion is in the form of the so-called Dunham expansion~\cite{Dunham_32}, which is used to describe a non-rigid diatomic molecule, whose interatomic distance increases when it rotates faster~\cite{huber2013molecular}. Therefore, as it has been shown in Refs.~\cite{esteve1992q,raychev1995quantum}, such a non-rigid rotor can be described by a quantum group  SO$_q$(3) with  $|q|=1$.

\begin{figure}
  \centering
  \includegraphics[width=0.9\linewidth]{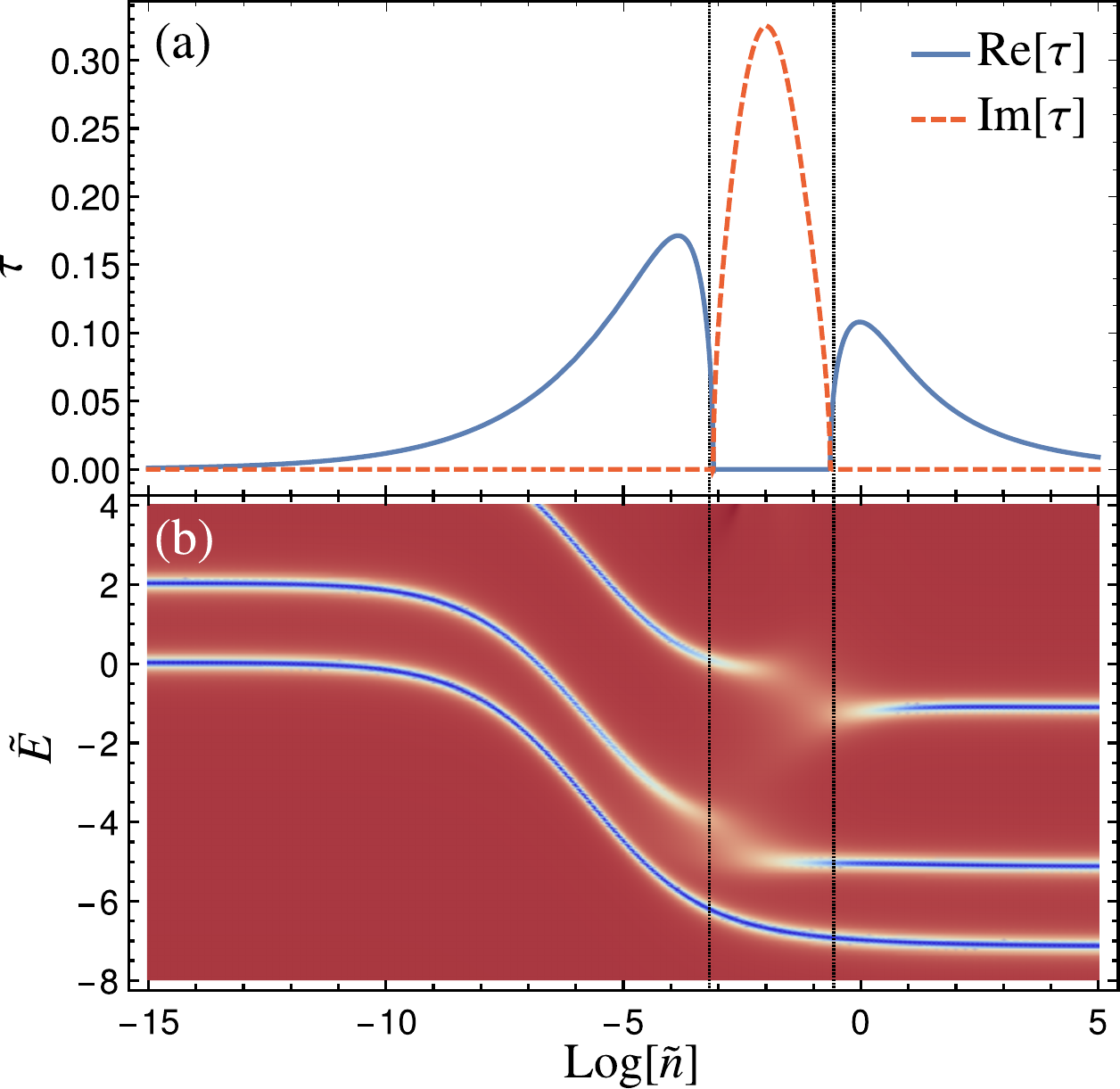}
 \caption{(a) The deformation parameter, $\tau$, as a function of the dimensionless bath density, $\tilde{n}$. (b) The spectral function of the angulon quasiparticle as a function of the dimensionless energy, $\tilde{E} = E/B$, and $\tilde{n}$,  obtained using the variational approach of Ref.~\cite{PhysRevX.6.011012}. The blue sharp peaks correspond to quasiparticle states, whereas yellowish blurred peaks show angulon instabilities, which match with the domain where $\tau$ is imaginary. The model parameters are the same as in Ref.~\cite{bighin2018diagrammatic}. See the text.}
 \label{tau}
\end{figure}

From the correspondence between the Dunham expansion and the quantum group, we deduce that when the deformation parameter $\tau$ is real, a rotor immersed in a many-particle bath manifests itself as a non-rigid rotor with a renormalized rotational constant, $B^\ast < B$. In the context of quantum impurity problems, the latter corresponds to the angulon quasiparticle~\cite{lemeshko2016quasiparticle, Lemeshko_2016_book}.  This can be seen from Fig.~\ref{tau}~(a), where the real values of $\tau$ correspond to the blue sharp peaks of  Fig.~\ref{tau}~(b).  As shown in the same figure, the deformation parameter $\tau$ can also assume imaginary values, which signals the breakdown of the quasiparticle picture. In fact, the latter corresponds to the so-called angulon instabilities~\cite{Lemeshko_2015,Cherepanov,Yakaboylu_2017,Yakaboylu_2017_b}, shown in Fig.~\ref{tau}~(b). We would like to emphasize that since the angulon instabilities have been already observed in the experiment~\cite{Cherepanov}, the imaginary values of the deformation parameter obtained within the quantum group approach correspond to a physical phenomenon.

As a final remark, we would like to note a striking analogy with knot theory where computation of a quantity within a classical group formalism is greatly simplified if one postulates a corresponding quantum group symmetry. More specifically, the Wilson loop observable in Chern-Simons theory assigns to a knot, combined with a representation of a classical group, a number depending on the level parameter~\cite{witten1989quantum}. This number is defined via path integration which is not completely satisfactory from the mathematical perspective. However, it can be rendered rigorously in two fashions. 

The first one consists of a perturbative expansion with respect to the level parameter. This leads to a very complicated universal series involving analogues of Feynman diagrams. Computed in a light-cone gauge~\cite{labastida1999chern}, this series is called the Kontsevich integral~\cite{kontsevich1993vassiliev,bar1995vassiliev}. This method has a great theoretical value, however, it is extremely difficult --  and often impossible -- to implement in practice~\cite{chmutov2012introduction}. The second method~\cite{reshetikhin1991invariants} can be  vaguely described as postulating that the knot is the world line of a particle obeying quantum group symmetry. As a result, the Wilson loop observable becomes accessible through a rather trivial product expression involving the deformation parameter $q$ which is related to the level parameter of the theory in a very explicit way. 

The mechanism of the drastic simplification is not completely understood neither in mathematics nor in physics. The discovery of such a phenomenon in the contex of quantum impurity problems indicates that it might be a general feature of complex quantum systems and thereby provide a general strategy for simplifying computations via quantizing symmetries.

Thus, under the hypothesis that quantum group is a hidden symmetry of a quantum impurity, we have shown that the effect of a many-particle environment on a rigid rotor can be seen as a deformation of the Lie group SO(3) to a quantum group, SO$_q$(3). We demonstrated that by evaluating the deformation parameter, $q$, from the closed-form perturbative expansion at weak coupling, one acquires access to the solutions at arbitrary coupling strengths through the quantum group formalism. We anticipate that the presented approach might be quite general and can be applied to, e.g., the polaron or spin impurity problems. Since quantum impurities represent elementary building blocks of many-particle systems, the approach based on quantum groups paves the way to uncover hidden symmetries in strongly correlated  matter and thereby drastically simplify its theoretical description.

\begin{acknowledgments}
We are grateful to A. Deuchert and I. Ganev for valuable discussions, and to G. Bighin for providing the numerical data from Ref.\cite{bighin2018diagrammatic}. E. Y. acknowledges financial support received from the People Programme (Marie Curie Actions) of the European Union's Seventh Framework Programme (FP7/2007-2013) under REA grant agreement No. [291734]. M. L. acknowledges support from the Austrian Science Fund (FWF), under project No. P29902-N27.
\end{acknowledgments}

\bibliography{yakaboylu_bibliography_ist.bib}

\end{document}